\documentclass[prb,a4paper,showpacs,twocolumn,superscriptaddress,longbibliography]{revtex4-1}

\usepackage{amssymb}
\usepackage{amsmath}
\usepackage{amsfonts}
\usepackage{graphicx}
\usepackage{bm}
\usepackage{color}
\usepackage{multirow}
\usepackage{natbib}
\usepackage{hyperref}
\usepackage[normalem]{ulem}

\DeclareMathOperator{\Tr}{Tr}

\DeclareMathOperator{\sgn}{sgn}

\DeclareMathOperator{\re}{Re}

\begin{document}

\title{Indirect exchange interaction between magnetic impurities in the two-dimensional topological insulator based on CdTe/HgTe/CdTe quantum wells}

\author{P. D. Kurilovich}

\affiliation{Moscow
Institute of Physics and Technology, 141700 Moscow, Russia}

\author{V. D. Kurilovich}

\affiliation{Moscow
Institute of Physics and Technology, 141700 Moscow, Russia}

\author{I. S.~Burmistrov}

\affiliation{L.D. Landau Institute for Theoretical Physics, Kosygina
  street 2, 117940 Moscow, Russia}
  
\affiliation{Moscow
Institute of Physics and Technology, 141700 Moscow, Russia}

\affiliation{Condensed-matter Physics Laboratory, National Research University Higher School of Economics, 101000 Moscow, Russia}

\begin{abstract}
We study indirect exchange interaction between magnetic impurities in the (001) CdTe/HgTe/CdTe symmetric quantum well. We consider low temperatures and the case of the chemical potential placed in the energy gap of the 2D quasiparticle spectrum. We find that the indirect exchange interaction is suppressed exponentially with the distance between magnetic impurities.  The presence of inversion asymmetry results  in oscillations of  the indirect exchange interaction with the distance and generates additional terms which are non-invariant under rotations in the (001) plane. The indirect exchange interaction matrix has complicated structure with some terms proportional to the sign of the energy gap.
 \end{abstract}


\maketitle

\section{Introduction}
\label{s1}

Theoretical prediction [\onlinecite{Kane-Mele},\onlinecite{BHZ}] and experimental observation [\onlinecite{Konig2007}] of 
the quantum Hall spin (QHS)  effect in a HgTe/CdTe quantum well (QW) triggered great interest to this two-dimensional (2D) topological insulator (TI) material [\onlinecite{Qi-Zhang},\onlinecite{Hasan-Kane}]. The existence of the QHS state with the perfect edge transport requires the presence of the time reversal symmetry. A global time reversal symmetry breaking perturbation, e.g. an external magnetic field, demolishes the QHS state. A local perturbation which breaks the time reversal symmetry does not destroy the QHS state but can affect the edge transport [\onlinecite{Maciejko2009},\onlinecite{Tanaka2011}]. 

A well-known example of the local time reversal symmetry breaking perturbation is a magnetic impurity in the classical limit. In the presence of a finite concentration of magnetic impurities the existence of the QHS state can be questioned. For example, if spins of magnetic impurities are ordered ferromagnetically then the QHS state will be suppressed due to the Zeeman splitting induced by a magnetization. 

For a small concentration of magnetic impurities a possible phase diagram (e.g. in the plane of temperature versus concentration) is determined by an indirect exchange interaction. In case of metals it is known as Ruderman-Kittel-Kasuya-Yosida (RKKY) interaction [\onlinecite{Ruderman-Kittel,Kasuya,Yosida}]. This interaction is long-ranged and oscillating in sign with the distance. The latter leads to the spin glass phase at low temperatures. Recently, RKKY interaction mediated by surface states of a three dimensional (3D) topological insulator attracted a lot of interest [\onlinecite{Zhang2009I,Ye2010,Garate-Franz,Biswas-Balatsky,Abanin-Pesin,Zhu2011,Efimkin-Galitski}]. It was predicted that the RKKY interaction can lead to ferromagnetic ordering of magnetic impurities and, thus, can open a gap in the spectrum of surface states. This effect was studied experimentally by angle resolved photoemission spectroscopy [\onlinecite{Chen2010,Wray2011,Xu2012}].

Studies of the indirect exchange interaction in semiconductors have been pioneered by Bloembergen and Rowland [\onlinecite{Bloembergen1955}]. It was shown that at zero temperature the presence of a finite gap between valence and conduction bands results in the exponential decay of the indirect exchange interaction with the distance provided the chemical potential lies within the gap. Such short-ranged ferromagnetic Heisenberg interaction between magnetic impurities results in the ferromagnetic state at low temperatures (see Ref. [\onlinecite{Korenblit1978},\onlinecite{Abrikosov1980}] for a review). 

The presence of strong spin-orbit coupling in a semiconductor complicates the form of the indirect exchange interaction. In the simplest case, in addition to the Heisenberg term the indirect exchange interaction involves also the magnetic pseudo-dipole interaction [\onlinecite{Kossut,Lewiner1980,Zarand}]. The latter prevents the system of magnetic impurities from ferromagnetic ordering and favors spin glass state. This was studied experimentally in details, for example, in diluted magnetic semiconductors Hg$_{1-x}$Mn$_x$Te and Cd$_{1-x}$Mn$_x$Te (see Ref. [\onlinecite{LyapilinTsidilkovski}] for a review). 

In this paper we study indirect exchange interaction between magnetic impurities  mediated by 2D electron and hole quasiparticles in the (001) CdTe/HgTe/CdTe symmetric QW.  We focus on the case of low temperatures and the chemical potential pinned within the energy gap of the 2D quasiparticle spectrum. We take into account the presence of inversion asymmetry in the QW [\onlinecite{Dai2008,Konig2008,Winkler2012,Weithofer-Recher,Tarasenko2015}]. We demonstrate that the indirect exchange interaction is suppressed exponentially with the distance between magnetic impurities. This is in accordance with a general expectations, since the exchange interaction is mediated by interband virtual transitions [\onlinecite{Abrikosov1980}]. We find that the presence of strong spin-orbit coupling in HgTe and CdTe semiconductors results in the following interesting features of the indirect exchange interaction:
\begin{itemize}
\item[(i)] In the absence of inversion asymmetry the interaction includes anisotropic XXZ Heisenberg interaction, magnetic pseudo-dipole interaction, and Dzyaloshinsky-Moriya interaction. 
The relative strengths of these terms depend on arrangement of magnetic impurities along the $z$ axis. In the case of magnetic impurities situated in the $x-y$ plane passing exactly through the middle of the QW, anisotropic XXZ Heisenberg interaction survives only. The sign of DM  interaction depends on the sign of the gap, i.e. the indirect exchange interaction distinguishes the trivial and topological insulators. Provided spins of 
magnetic impurities are polarized in $z$ direction, the indirect exchange interaction is of antiferromagnetic sign. 
\item[(ii)] The presence of inversion asymmetry makes the indirect exchange interaction oscillating with the distance and generates additional terms which are non-invariant under rotations in the $x-y$ plane. Some of these terms are proportional to the sign of the gap. Provided magnetic impurities are placed exactly in the $x-y$ plane passing through the middle of the QW the resulting indirect exchange interaction is compatible with the $D_{2d}$ symmetry of the system. In this case there is no dependence on the sign of the gap. If spins of magnetic impurities are polarized in $z$ direction, the sign of indirect exchange interaction depends on the distance between them. 
\end{itemize}

The outline of the paper is as follows.  In Sec. \ref{Sec:BHZ} we remind a reader the Hamiltonian for  2D electron and hole states in the (001) CdTe/HgTe/CdTe QW and derive the effective 2D Hamiltonian for the magnetic impurity. The indirect exchange interaction is derived in Sec. \ref{Sec:IEI}. The discussion of our result and conclusions are presented in Sec. \ref{Sec:DisConc}. Technical details of derivation of the indirect exchange interaction are given in Appendix \ref{App1}. Appendix \ref{App2} contains discussion of the single-spin anisotropy Hamiltonian. 

\section{The effective 2D Hamiltonian for a magnetic impurity \label{Sec:BHZ}}

The 2D electron and hole states in a (001) CdTe/HgTe/CdTe QW are described by the effective $4\times 4$ Bernevig-Hughes-Zhang (BHZ) Hamiltonian [\onlinecite{BHZ}]. This Hamiltonian can be written with the help of symmetry considerations or, alternatively, derived by means of the $\bm{k}\cdot\bm{p}$ method. In order to evaluate the indirect exchange interaction between magnetic impurities it is necessary to derive the corresponding $4\times 4$ Hamiltonian for a magnetic impurity.

\subsection{BHZ Hamiltonian with bulk and surface induced inversion asymmetry}

To set notations, we review the $\bm{k}\cdot\bm{p}$ method of description of the electronic band structure in the bulk of crystals with zinc blend structure (see Ref. [\onlinecite{BirPikus}] for details). Electronic bands in the $\Gamma$ point of these crystals can be classified according to the $T_d$ double group representations: $\Gamma_6$, $\Gamma_7$ and $\Gamma_8$. The effective  
$\bm{k}\cdot\bm{p}$ Hamiltonian that describes coupling between the $\Gamma_6$ and $\Gamma_8$ bands while taking into account the other bands as perturbations is known as $6\times 6$ Kane Hamiltonian [\onlinecite{Kane1957}]:
\begin{gather}\label{Kane}
{H}_{\rm Kane}(\bm{k})=\begin{pmatrix} H_c(\bm{k}) & T(\bm{k})\\
T^{\dagger}(\bm{k}) & H_v(\bm{k})
\end{pmatrix} .
\end{gather}
Here the $\Gamma_6$ band is described by $2\times 2$ diagonal Hamiltonian
\begin{equation}
H_c(\bm{k})=E_c+\frac{k^2}{2m_*},
\end{equation}
where $m^*$ and $E_c$ are the effective mass and the energy of the bottom of the conduction band, respectively. The $4\times4$ Hamiltonian 
\begin{equation}
H_v(\bm{k})=E_v-(\gamma_1+\frac{5}{2}\gamma_2)\frac{k^2}{2m_0}+\frac{\gamma_2}{m_0}(\bm{k}\cdot \bm{J}_{3/2})^2
\end{equation}
describes the $\Gamma_8$ band. Here $E_v$ is the energy of the top of the valence band, $m_0$ is the  electron mass, $\gamma_1$ and $\gamma_2$ are the Luttinger parameters [\onlinecite{Luttinger1956}], $\bm{J}_{3/2}$ denotes the spin 3/2 operator. At non-zero values of 3D wave vector $\bm{k}$ the $\Gamma_6$ and $\Gamma_8$ bands are coupled by the following $2\times 4$ matrix:
\begin{equation}
T(\bm{k})=\begin{pmatrix}
- \frac{1}{\sqrt{2}}Pk_{+} & \sqrt{\frac{2}{3}}Pk_z & \frac{1}{\sqrt{6}}Pk_{-} & 0\\         
 0 & -\frac{1}{\sqrt{6}}Pk_{+} &    \sqrt{\frac{2}{3}}Pk_z & \frac{1}{\sqrt{2}}Pk_{-} 
\end{pmatrix} .
\end{equation}
Here $P$ is the Kane matrix element and $k_\pm = k_x\pm i k_y$. The $z$ axis is the QW growth direction $[001]$ and the in-plane axes $x$ and $y$ are parallel to [100] and [010] directions. The Hamiltonian above is written in the standard basis:
$|\Gamma_6,+1/2 \rangle$, 
$|\Gamma_6,-1/2 \rangle$, 
$|\Gamma_8,+3/2 \rangle$, 
$|\Gamma_8,+1/2 \rangle$, 
$|\Gamma_8,-1/2 \rangle$, 
$|\Gamma_8,-3/2 \rangle$.~\footnote{
We remind that the states are enumerated as follows. Both $\Gamma_6$ and $\Gamma_8$ representations are also twofold  and fourfold representations of the full rotation group, respectively. Thus, one can characterize their basis functions with the value of corresponding angular momentum and its projection in a sense that these functions transform as the functions with this angular momentum under the rotations from the $T_d$ group. The value of angular momentum projection is given next to the name of the representation.}

In the (001) CdTe/HgTe/CdTe QW the spatial quantization happens. The solutions of the corresponding 
Schr\"odinger equation with the Hamiltonian \eqref{Kane} with $k_{x}=k_y=0$ can be written as
\begin{align}
|E_i \pm\rangle & = f^{(i)}_{1,2}(z)|\Gamma_6,\pm 1/2\rangle + f^{(i)}_{4,5}(z)|\Gamma_8,\pm 1/2\rangle, \notag  \\
|H_i \pm \rangle & = f^{(i)}_{3,6}(z)|\Gamma_8,\pm 3/2\rangle, \label{eq:states} \\
|L_i \pm \rangle & = g^{(i)}_{1,2}(z)|\Gamma_6,\pm1/2\rangle + g^{(i)}_{4,5}(z)|\Gamma_8,\pm1/2\rangle ,\notag
\end{align}
where $i=1,2,\dots$ stands for the quantum number of a level of spatial quantization. At low temperatures the lowest level of the spatial quantization is important only. The functions 
$f^{(1)}_{1,2,3,6}(z)$  and $g^{(1)}_{4,5}$ are symmetric under inversion $z \rightarrow -z$, while $f^{(1)}_{4,5}$ and $g^{(1)}_{1,2}$ are antisymmetric. \footnote{For further details on the function $f^{(i)}_j(z)$ and $g^{(i)}_j(z)$ see Supporting Online Material for Ref. [\protect\onlinecite{BHZ}]} Since the energy of the light hole states $|L_1 \pm \rangle$ is well above the energies of the electron $|E_1 \pm \rangle$ and heavy hole state $|H_1 \pm \rangle$ we can project the $6 \times 6$ Hamiltonian \eqref{Kane} onto the low energy subspace  ($\vert E_1,+\rangle$, $\vert H_1,+\rangle$, $\vert E_1,-\rangle$, $\vert H_1,-\rangle$). The result is known as the BHZ Hamiltonian:
\begin{equation} 
H_{\rm BHZ}= \varepsilon(k) +
\begin{pmatrix} 
M(k)&Ak_{+}&0&0\\ 
Ak_{-}&-M(k)&0&0\\ 
0&0&M(k)&-Ak_{-}\\ 
0&0&-Ak_{+}&-M(k)\\ 
\end{pmatrix}  ,
\label{BHZ-ham}
\end{equation} 
where 
\begin{equation}
\varepsilon(k)=C-D(k_x^2+k_y^2), \quad 
M(k)=M-B(k_x^2+k_y^2) .
\end{equation} 
The material parameters $A$, $B$, $C$, $D$ and $M$ depends on the width $d$ of the CdTe/HgTe/CdTe QW. In what follows we will measure all energies with respect to a value of $C$. Values of the other four parameters for several values of the QW width can be found in Table I of Ref. [\onlinecite{Qi-Zhang}]. 
The parameter $M$ is positive (negative) for $d<d_c$ ($d>d_c$). The critical width $d_c\approx 6.3$ nm corresponds to the quantum phase transition between 2D trivial insulator and topological insulator. Also, we note that in the vicinity of the critical width both parameters $B$ and $D$ are negative and $|B|>|D|$.

The Hamiltonian \eqref{BHZ-ham} is invariant under rotation in the $x-y$ plane and, consequently, does not much sensitive to details of the crystal symmetry of the QW.  The time-reversal symmetry allows to add the following term to the BHZ Hamiltonian:
\begin{equation} 
H_{\rm ia}= 
\begin{pmatrix} 
0&0&0&\Delta\\ 
0&0&-\Delta&0\\ 
0&-\Delta&0&0\\ 
\Delta&0&0&0\\ 
\end{pmatrix} .
\label{H-ia}
\end{equation}
This term breaks the rotational invariance in the $x-y$ plane. Rotation of the system on angle $\alpha$ around the $z$-axis transforms $\Delta \to \Delta \exp(2i\alpha)$ in the upper-right $2\times 2$ block of the Hamiltonian \eqref{H-ia} and $\Delta \to \Delta \exp(-2i\alpha)$ in the lower-left $2\times 2$ block. A nonzero value of $\Delta$ can exist due to the bulk inversion asymmetry: an inversion element in the $T_d$ group is absent
[\onlinecite{Konig2008,Dai2008,Winkler2012,Weithofer-Recher}]. The other reason of nonzero value of $\Delta$ is the interface inversion asymmetry. It is associated with the natural (atomistic) non-equivalence between the top and bottom interfaces of the symmetric (001) CdTe/HgTe/CdTe QW. This inequivalence results in a $D_{2d}$ symmetry of such QW [\onlinecite{Tarasenko2015}]. The atomistic calculations of Ref. [\onlinecite{Tarasenko2015}] demonstrates that the interface inversion asymmetry induces contribution to $\Delta$ of the order of $5 - 10$ meV for QWs with widths close to $d_c$. The contribution to $\Delta$ due to the bulk inversion asymmetry is estimated to be several times smaller [\onlinecite{Konig2008},\onlinecite{Winkler2012}].  Recent experiments [\onlinecite{Minkov2013,Minkov2016}] revealed the presence of large splitting $\Delta$ of electron and heavy hole states indeed. We emphasize that both bulk and interface inversion asymmetry leads to Hamiltonian \eqref{H-ia}. Following Ref. [\onlinecite{Durnev2016}], we will use value of $\Delta=5$ meV for numerical estimates below.

\subsection{Magnetic impurity}

Microscopically, interaction between the spin $\bm{S}$ of a magnetic impurity  and the electron and hole spins is described by a standard exchange Hamiltonian $j(\bm{r}) \bm{S}\cdot\bm{\sigma}$ where $j(\bm{r})$ is a short-ranged potential.  In the bulk of the crystal with $T_d$ symmetry this Hamiltonian projected to the $\Gamma_6$ and $\Gamma_8$ states becomes the following $6\times 6$ matrix (see for example, Ref. [\onlinecite{Winkler}]:
\begin{equation}
V_{\rm imp} =
\begin{pmatrix}
2{j}_6(\bm{r})\bm{J}_{1/2}\cdot \bm{S} & 0\\
0 & \frac{2}{3}j_8(\bm{r}) \bm{J}_{3/2} \cdot \bm{S}
\end{pmatrix} .
\label{Vimp1}
\end{equation}
Here $\bm{J}_{S}$ is the spin  $S$ operator. The effective potentials $j_6(\bm{r})$ and $j_8(\bm{r})$ are proportional to the microscopic potential $j(\bm{r})$ and the effective $g$-factors for the conduction and valence bands, respectively [\onlinecite{Winkler}]. In general, nontrivial terms of higher order in $\bm{S}$ describing interaction between a magnetic impurity  and the electron and hole states are possible. Since we are interested in contribution to the indirect exchange interaction of the lowest order in spin operators we shall not consider such terms in the present paper.

For description of a magnetic impurity in the CdTe/HgTe/CdTe QW we need to project Hamiltonian \eqref{Vimp1} onto the low energy space of electron and heavy hole states: $|E_1,+\rangle$, $|H_1,+\rangle$, $|E_1,-\rangle$, and $|H_1,-\rangle$. We assume that the range of an impurity potential is much larger than the atomic one but much shorter than the scale at which the envelope functions $f_i^{(1)}$ changes, i.e.  we assume that $J_6(\bm{r})=\alpha \delta(\bm{r}-\bm{r}_0)$ and $J_8(\bm{r})=\beta \delta(\bm{r}-\bm{r}_0)$ where $\bm{r_0}$ denotes the position of the magnetic impurity. Then to the lowest order in $\alpha$ and $\beta$ the effective $4\times4$ Hamiltonian for a magnetic impurity in the CdTe/HgTe/CdTe QW assumes the following  form:
\begin{equation}
\mathcal{V}_{\rm imp} = \mathcal{J} \delta(x-x_0) \delta(y-y_0) ,
\end{equation}
where
\begin{equation}
\mathcal{J} =
\begin{pmatrix}
J_1 S_z & -iJ_0 S_+ & J_{m} S_{-} &0\\
iJ_0S_{-} & J_2 S_z &0 &0\\
J_{m} S_{+} &0 & -J_1 S_z & - iJ_0 S_{-}\\
0 & 0 &  iJ_0 S_+ & -J_2 S_z   
\end{pmatrix} .
\label{eq:Jmat}
\end{equation}
Here the real parameters $J_0, J_1, J_2$ and $J_m$ are defined in terms of the envelope functions:
\begin{equation}
\begin{split}
J_0 & = \frac{i \beta}{\sqrt{3}} f_3^{(1)}(z_0)\overline{f^{(1)}_4(z_0)}, \\
J_1 & = \alpha |f_1^{(1)}(z_0)|^2 + \frac{\beta}{3} |f_4^{(1)}(z_0)|^2,  \\  
J_2 & = \beta |f_3^{(1)}(z_0)|^2, \\
J_m & = J_1+J_0^2/J_2.
\end{split}
\label{eqJ}
\end{equation}
Here we use the following properties of envelope functions: (i) $f_1^{(1)}=f_2^{(1)}$, $f_3^{(1)}=f_6^{(1)}$ and $f_4^{(1)}=f_5^{(1)}$; (ii) the functions $f_{1,2,3,6}^{(1)}$ are real, $\overline{f^{(1)}_{1,2,3,6}}=f^{(1)}_{1,2,3,6}$, and $f^{(1)}_{4,5}$ are imaginary, $\overline{f^{(1)}_{4,5}}=-f^{(1)}_{4,5}$. 
It is worthwhile to mention that a magnetic impurity couples  the electron subbands $|E_1,+\rangle$ and $|E_1,-\rangle$ related by time-reversal symmetry. There is no coupling between the subbands $|E_1,+\rangle$ and $|H_1,-\rangle$, or $|H_1,+\rangle$ and $|E_1,-\rangle$, or $|H_1,+\rangle$ and $|H_1,-\rangle$.
This can be explained as follows: the electron (heavy hole) states have the projection of angular momentum $\pm {1}/{2}$ ($\pm {3}/{2}$) whereas a magnetic impurity can flip the electron spin and change its projection by $\pm 1$. 

Finally, we mention that if a magnetic impurity is situated at the center of the symmetric QW, $z_0=0$, the antisymmetric function $f_4$ at the impurity position vanishes. Hence, one finds $J_0=0$ and $J_m=J_1$.

\section{Indirect exchange interaction\label{Sec:IEI}}

To the second order in $\mathcal{J}$ the indirect exchange interaction is given by a standard spin-susceptibility-type diagram. The corresponding effective Hamiltonian describing interaction of two magnetic impurities situated at points $\bm{r_A}=\{\bm{R_A},z_A\}$ and $\bm{r_B}=\{\bm{R_B},z_B\}$ can be written as
\begin{equation}
{H}_{\rm IEI} = T\sum_{\varepsilon_n} \Tr  \mathcal{J}^A \mathcal{G}(i\varepsilon_n, \bm{R_A}, \bm{R_B}) 
 \mathcal{J}^B \mathcal{G}(i\varepsilon_n, \bm{R_B}, \bm{R_A})  .
 \label{H-IEI}
\end{equation}
Here, $\varepsilon_n=\pi T (2n+1)$ denotes the fermionic Matsubara frequencies. The Matsubara Green's function corresponding to the Hamiltonian 
$\mathcal{H}=H_{\rm BHZ}+ H_{\rm ia}$ is given as
\begin{equation}
\begin{split}
 \mathcal{G}(i\varepsilon_n, \bm{R_A}, \bm{R_B}) = 
 \int \frac{d^2\bm{k}}{(2\pi)^2} e^{i \bm{k} \bm{R}} \mathcal{G}(i\varepsilon_n,\bm{k}),  \\
\mathcal{G}(i\varepsilon_n,\bm{k}) = \Bigl [i\varepsilon_n +\mu - \mathcal{H} \Bigr ]^{-1},
 \end{split}
\label{eq:full-hamGF}
\end{equation}
where $\bm{R} = \bm{R_A}-\bm{R_B}$ and $\mu$ denotes the chemical potential.
The superscript $A$ ($B$) in $\mathcal{J}^A$ ($\mathcal{J}^B$) indicates that the matrix \eqref{eq:Jmat} is evaluated at the position $z_A$ ($z_B$). 

To proceed further we introduce convenient unites. At first, we introduce the characteristic length scale ($a$) and energy scale ($\mathcal{E}$) in the problem:
\begin{equation}
a=\sqrt{B^2-D^2}/{A}, \quad \mathcal{E} ={A^2}/\sqrt{B^2-D^2} .
\end{equation}
Secondly, we introduce the following dimensionless parameters:
\begin{equation}
m = \frac{M}{\mathcal{E}}, \quad \cosh{\chi}=-\frac{B}{\sqrt{B^2-D^2}}, \quad \gamma = \frac{\Delta}{|m| \mathcal{E}} .
\end{equation}
The numerical estimates of these parameters obtained with the help of Table I of Ref. [\onlinecite{Qi-Zhang}] and for the value $\Delta\approx 5$ meV [\onlinecite{Tarasenko2015},\onlinecite{Durnev2016}] are summarized in Table \ref{Tab}. We note that the dimensionless gap is small, $|m| \ll 1$.
Finally, we define dimensionless vectors $\bm{\kappa} = \bm{k} a/|m|$ and $\bm{\rho} = \bm{R} |m|/a$.

Before evaluation of Eq. \eqref{H-IEI} it is convenient to diagonalize the Green's function:
\begin{equation}
\mathcal{G}(i\varepsilon_n,\bm{k}) = \mathcal{R}(\bm{\kappa}) \hat{\mathcal{G}}(i\varepsilon_n,\bm{\kappa}) 
\mathcal{R}^{-1}(\bm{\kappa}) ,
\end{equation}
where
\begin{equation}
\hat{\mathcal{G}}^{-1} = i\varepsilon_n +\mu - |m| \mathcal{E} 
\begin{pmatrix} 
\epsilon_1(\bm{\kappa})& 0 & 0 & 0\\
0 & \epsilon_2(\bm{\kappa}) & 0 & 0\\
0 & 0 & \epsilon_3(\bm{\kappa}) & 0\\
0 & 0 & 0 & \epsilon_4(\bm{\kappa})  
\end{pmatrix}.
\label{eq:Gdiag}
\end{equation}
The energy spectrum for the Hamiltonian $\mathcal{H}$ is given as 
\begin{equation}
\begin{split}
\epsilon_{1,3} & =  -  \kappa^2  |m| \sinh \chi+ \sgn m \sqrt{(\kappa \pm \gamma)^2+b^2(\kappa)}
  ,\\
\epsilon_{2,4} & = -  \kappa^2  |m| \sinh \chi - \sgn m \sqrt{(\kappa \pm \gamma)^2+b^2(\kappa)} 
 ,
\end{split}
\label{eq:sp}
\end{equation}
where $b(\kappa^2)=1+\kappa^2 m \cosh \chi$.
It is worthwhile to mention that in the presence of inversion asymmetry (i) the electron and heavy hole subbands become non-degenerate at $\mathbf{k}\neq 0$ and (ii) the spectrum remains isotropic in the $x$ -- $y$ plane. The matrix $\mathcal{R}$ is given as
\begin{equation}
\mathcal{R} = \begin{pmatrix}
\frac{{\cal{M}}_{++}}{\gamma+\kappa} & \frac{{\cal{M}}_{-+}}{\gamma+\kappa}  & \frac{{\cal{M}}_{+-}}{\gamma-\kappa}  & \frac{{\cal{M}}_{--}}{\gamma-\kappa} \\
e^{-i\theta} & e^{-i\theta} & -e^{-i\theta} & -e^{-i\theta}\\
\frac{-e^{-i\theta}{\cal{M}}_{++}}{\gamma+\kappa} & \frac{-e^{-i\theta}{\cal{M}}_{-+}}{\gamma+\kappa}  & \frac{e^{-i\theta}{\cal{M}}_{+-}}{\gamma-\kappa}  & \frac{e^{-i\theta}{\cal{M}}_{--}}{\gamma-\kappa} \\
1 & 1 & 1 & 1 
\end{pmatrix} , 
\end{equation}
where $\theta$ stands for the angle between $\bm{\kappa}$ and $x$ axis, and 
\begin{equation}
{\cal{M}}_{pq}=\sgn m \:\Bigl (b(\kappa^2)+p \sqrt{(\kappa +q  \gamma)^2+b^2(\kappa)}\Bigr ) .
\end{equation} 
At zero temperature and for the chemical potential pinned to the gap, $|\mu| < |M|$, the summation over Matsubara frequencies in Eq. \eqref{H-IEI} can be easily performed. Then we find

\begin{equation}
{H}_{\rm IEI} = \mathcal{J}^A_{\alpha\beta} \mathcal{J}^B_{\gamma\delta} \Pi_{\beta\gamma,\delta\alpha}(\bm{R_A},\bm{R_B}), 
\end{equation}
where the polarization operator is given as follows
\begin{align}
\Pi_{\beta\gamma,\delta\alpha}&  = - \frac{|m|^3}{a^4 \mathcal{E}} \int\limits_0^\infty dt \int \frac{d^2\bm{\kappa_1}d^2\bm{\kappa_2}}{(2\pi)^4} e^{i \bm{\rho} (\bm{\kappa_1}-\bm{\kappa_2})}
\notag\\
  \times & \sum_{j=1,3} \sum_{l=2,4} e^{-t \sgn m [\epsilon_j(\kappa_1)-\epsilon_l(\kappa_2) ]} \Biggl \{\mathcal{R}_{\beta l}(-\bm{\kappa_2})
  \notag \\
  \times & 
\mathcal{R}^{-1}_{l\gamma}(-\bm{\kappa_2})
\mathcal{R}_{\delta j}(-\bm{\kappa_1})
\mathcal{R}^{-1}_{j \alpha}(-\bm{\kappa_1})
\notag \\
+ & 
\mathcal{R}_{\beta j}(\bm{\kappa_1})
\mathcal{R}^{-1}_{j\gamma}(\bm{\kappa_1})
\mathcal{R}_{\delta l}(\bm{\kappa_2})
\mathcal{R}^{-1}_{l\alpha}(\bm{\kappa_2})
\Biggr \} .
\label{eq:Pi}
\end{align}
The polarization operator at small distances $\rho \lesssim 1$ is not universal. In this case its evaluation requires knowledge of the quasiparticle spectrum in the whole Brillouin zone. Therefore, in what follows we focus on large distance regime $\rho \gg 1$ (or $R \gg a/|m|$). In this limit the integrals in Eq. \eqref{eq:Pi} can be evaluated with the help of the saddle point approximation. Then the indirect exchange interaction becomes (see details in Appendix \ref{App1}):
\begin{equation}
{H}_{\rm IEI} = \sum_{a,b=x,y,z} K_{ab} S^A_a S^B_b , 
\label{eq:IEI-1}
\end{equation}
where
\begin{align}
K_{xx}  & =
J_m^A J_m^B \Bigl [ F(R) n_x^2 +  F_c(R) n_y^2 \Bigr ] - 4 J_0^A J_0^B F_c(R) n_x^2  \notag\\
& - 2 \sgn M \,(J_0^AJ_m^B+J_m^AJ_0^B) F_s(R)   n_x n_y ,  \notag \\
K_{xy} & = \Bigl [ J_m^A J_m^B \Bigl ( F_c(R)-F(R)\Bigr )  - 4 J_0^A J_0^B F_c(R) \Bigr  ] n_x n_y \notag\\
& - 2 \sgn M\, (J_0^AJ_m^B n_x^2 +J_m^AJ_0^B n_y^2)F_s(R), \notag \\
K_{xz} &  = 2 \sgn M\, J_0^A J_z^B F_c(R) n_x +  J_m^A J_z^B F_s(R) n_y ,\notag \\
K_{zz} & =  J_z^A J_z^B F_c(R) . 
\label{eq:IEI-2}
\end{align}
Here we introduce the two-dimensional unit vector $\bm{n}=\bm{R}/R$ and $J_z = J_1+J_2$.
The elements $K_{yy}$, $K_{yx}$ and $K_{yz}$ can be found from $K_{xx}$, $K_{xy}$ and $K_{xz}$ by interchange of $n_x$ and $n_y$. The elements $K_{zx}$ and $K_{zy}$ are equal to  $-K_{xz}$ and $-K_{yz}$ with superscripts $A$ and $B$ interchanged. The three functions $F(R)$, $F_c(R)$, and $F_s(R)$ are given as
\begin{equation}
F(R) = \frac{|m|^3 (1+\gamma^2)^{1/2}}{(2\pi)^{3/2} \mathcal{E} a^4} \left (\frac{\lambda_1}{R} \right )^{3/2}
e^{-R/\lambda_1},
\label{eq:IEI-3-0}
\end{equation}
and
\begin{equation}
\begin{split}
F_c(R)  & = F(R) 
\cos\bigl (R/\lambda_2 - \arctan \gamma\bigr ) ,
\\
F_s(R) 
 & = F(R)  \sin\bigl (R/\lambda_2 - \arctan \gamma\bigr ) .
\end{split} 
\label{eq:IEI-3}
\end{equation}
Here we introduce the following two characteristic length scales:
\begin{equation}
\begin{split}
\lambda_1 & = \frac{a}{2|m|} \bigl [1+(1-\gamma^2)m \cosh\chi \bigr ], \\
 \lambda_2 & = \frac{a}{2|m| \gamma} \bigl [1+2m \cosh\chi \bigr ] .
\end{split}
\label{eq:IEI-4}
\end{equation} 
The asymptotic expressions \eqref{eq:IEI-3} for the functions $F$, $F_c$, and $F_s$ are valid provided 
$R\gg \lambda_1$, $|m|\ll 1$, and $\gamma < 1$. Equations \eqref{eq:IEI-1}-\eqref{eq:IEI-4} constitute the main result of the present paper.

\begin{table}[t]
\caption{Parameters of the Hamiltonian $\mathcal{H}=H_{\rm BHZ}+H_{\rm ia}$ for two values of the QW width.}
\label{Tab}
\begin{tabular}{|c|c|c|c|c|c|c|c|}
\hline
$d$, $nm$ & $a$, $nm$  & $\mathcal{E}$, eV & $m$ & $\gamma$ & $\chi$ & $\lambda_1$, $nm$  &  $\lambda_2$, $nm$  \\ \hline
5.5 & 0.94  & 0.42 & 0.022 & 0.55 & -0.77 & 22 & 41\\ \hline
7.0 & 1.29  & 0.28 & -0.029 & 0.63 & -0.95 & 22 & 32\\ \hline
\end{tabular}
\end{table}

The indirect exchange interaction \eqref{eq:IEI-1} depends on the sign of the gap $M$, i.e. magnetic impurities interact differently in the trivial and topological phases. The finite value of the inversion asymmetry splitting $\gamma$ induces oscillations of the indirect exchange interaction with the distance. The estimates for $\lambda_1$ and $\lambda_2$ are presented in Table \ref{Tab}. As one can see, both length scales are large in comparison with $a$, $\lambda_2>\lambda_1\gg a$. Such oscillations in case of minima of the conduction and valence bands at non-zero $\bm{k}$ have been predicted long ago [\onlinecite{Abrikosov1980}]. Contrary to the previous expectations, in our case the indirect exchange interaction oscillates on distances much larger than the atomic one (see Table \ref{Tab}). This occurs  due to small value of the inversion asymmetry splitting $\Delta$. 

As we mentioned above, in the presence of nonzero $\Delta$ the Hamiltonian $\mathcal{H}$ is not invariant with respect to rotations around the $z$ axis. This holds for the indirect exchange interaction also. For $\gamma\neq 0$  $H_{\rm IEI}$ is not invariant with respect to rotations around the $z$ axis. If both magnetic impurities are situated in the plane at the center of the QW, i.e. $z_A=z_B=0$, the indirect exchange interaction is drastically simplified in virtue of the following condition: $J_0^A=J_0^B=0$. In this case, the form of the indirect exchange interaction is compatible with the $D_{2d}$ symmetry and is independent of the sign of the gap. 
 
In the absence of the inversion asymmetry, $\gamma=0$, the function $F_s$ vanishes whereas the functions $F$ and $F_c$ coincide, $F_c=F$. Then the indirect exchange interaction acquires the following form ($\bm{S}_\|=\{S_x,S_y\}$):
\begin{gather}
\, \hspace{-.8cm} {H}_{\rm IEI}  =\Biggl [ J_m^A J_m^A  \bigl ( \bm{S}_\|^A \cdot \bm{S}_\|^B \bigr )-
4 J_0^A J_0^B \bigl ( \bm{S}_\|^A \cdot \bm{n} \bigl ) \bigr (\bm{S}_\|^B \cdot \bm{n}\bigr )
\notag \\
+ 2 \sgn M  \Bigl ( J_0^A J_z^B \bigl ( \bm{S}_\|^A \cdot \bm{n} \bigl ) S_z^B - J_z^A J_0^B
S_z^A \bigl ( \bm{S}_\|^B \cdot \bm{n} \bigl )\Bigr )\notag \\
+  J_z^A J_z^B S_z^A S_z^B   \Biggr ] F(R) ,
\label{eq:IEI-5}
\end{gather}
where the function $F(R)$ is given by Eq. \eqref{eq:IEI-3-0} with $\gamma=0$.
As one can see, in this case the indirect exchange interaction includes anisotropic antiferromagnetic XXZ Heisenberg interaction, magnetic pseudo-dipole interaction, and DM interaction. The sign of the latter depends on the sign of the gap $M$.

It is worthwhile to mention that the spin structure of  Eq. \eqref{eq:IEI-5} resembles the indirect exchange  interaction mediated by gapped surface states of a 3D TI with a chemical potential within the gap [\onlinecite{Ye2010},\onlinecite{Abanin-Pesin},\onlinecite{Efimkin-Galitski}]. However, there are a number of important differences. (i) For a 3D TI the sign of indirect exchange interaction for spins aligned perpendicular to the surface (along the $z$ axis) is ferromagnetic whereas in Eq. \eqref{eq:IEI-5}  it is antiferromagnetic. (ii) For a surface of a 3D TI the DM term appears only when the chemical potential lies outside the gap whereas in our case it exists for the chemical potential pinned to the gap. (iii) In the case of a 3D TI the magnetic pseudo-dipole interaction is the most relevant term at large distances between magnetic impurities whereas in Eq. \eqref{eq:IEI-5} all terms behave in the same way with the distance.

Finally, we mention that it is possible to derive the large distance asymptote of the indirect exchange interaction for a more general Hamiltonian in which $b$, $\gamma$ and $\chi$ are arbitrary functions of $\kappa^2$ of the order of unity which change slowly between $\kappa=0$ and $\kappa \sim 1$ (see Appendix \ref{App1}).

\section{Discussion and conclusions\label{Sec:DisConc}}

Our derivation of the indirect exchange interaction \eqref{eq:IEI-1}-\eqref{eq:IEI-4} was done at zero temperature. In this case the interaction between magnetic impurities is insensitive to the position of the chemical potential within the gap, i.e. for $|\mu|<|M|$. At finite temperature $T \ll |M|$ this is not the case. However, our results \eqref{eq:IEI-3-0} are applicable at distances $R$ which satisfy the following inequality:
\begin{equation}
\frac{\mathcal{E} |M|}{T^2} \min\left \{1,\frac{T(1-\mu^2/M^2)}{|\mu|}\right\}\gg \frac{R}{a} \gg \frac{\mathcal{E}}{|M|} .
\end{equation}

The elements of the off-diagonal blocks of the Hamiltonian \eqref{H-ia} contain also terms which are linear in $\bm{k}$ [\onlinecite{Liu-Zhang},\onlinecite{Pikulin2014},\onlinecite{Durnev2016}]: 
\begin{equation}
H_{ia} \to \begin{pmatrix}
0 & 0 & \Delta_+ k_+ & \Delta \\
0 & 0 & -\Delta & \Delta_- k_- \\
\Delta_+ k_- & -\Delta & 0 & 0\\
\Delta & \Delta_- k_+ & 0 & 0
\end{pmatrix} .
\end{equation} 
In the presence of nonzero values of $\Delta_+$ and $\Delta_-$ linear in $\kappa$ term, $(\Delta_++\Delta_-)\kappa/A$ should be added to the function $b(\kappa^2)$ (see Eq. \eqref{eq:sp}). Using the estimates from Ref. [\onlinecite{Liu-Zhang}] we find that $(\Delta_++\Delta_-)/A$ is of the order of $10^{-2}$. This smallness justifies our analysis in which such terms are neglected.

The indirect exchange interaction \eqref{eq:IEI-1}-\eqref{eq:IEI-4} is computed to the lowest non-trivial order in the coupling between a magnetic impurity and the electron (E1) and heavy hole (H1) subbands in CdTe/HgTe/CdTe QW. A standard condition of validity of the perturbation theory implies in our case the following inequality:
\begin{equation}
(\max\{J_z,J_0,J_m\})^2 F(R) \ll |M| .
\label{eq:cond1}
\end{equation}
For an estimate we take the distance between impurities to be $R\sim \lambda_1$ which is the minimal distance at which Eqs. \eqref{eq:IEI-3} are valid. Then Eq. \eqref{eq:cond1} can be rewritten as follows:
\begin{equation}
|M| \max\{|\alpha|,|\beta|\}/(A^2 d) \ll 1. 
\label{eq:cond2}
\end{equation}
Taking $\max\{|\alpha|,|\beta|\} \approx 0.3 $ eV$\cdot$nm$^3$ [\onlinecite{Furdyna}] and using values of the parameters from Table \ref{Tab}, we find that the left hand side of inequality \eqref{eq:cond2} is of the order of $10^{-3}$.  This implies that the perturbation theory is well justified. It is worthwhile to mention that Eq. \eqref{eq:cond2} has a transparent meaning since combination $|M|/(A^2 d)$ is the 3D density of states at $k=0$ (for $\gamma=0$).

The decaying length of IEI between the magnetic impurities in the bulk of 3D CdTe crystal can be estimated as $\lambda_{\mathrm{bulk}}\sim 0.1 \div 1$ nm [\onlinecite{Bloembergen1955}]. We emphasize that
$\lambda_{\mathrm{bulk}} \ll \lambda_{1}$, i.e. IEI mediated by  2D states in the CdTe/HgTe/CdTe QW is much more long ranged.

At temperatures $T \gg T_* \sim (\max\{J_z,J_0,J_m\})^2 F(R\sim\lambda_1)$ a diluted system of magnetic impurities, $n_{\rm imp} \lesssim \lambda_1^{-2}$, behaves as the system of independent spins.  Since $T_*$ can be estimated to be of the order of $10^{-3}\div 10^{-4}$ K, magnetic impurities with concentrations $n_{\rm imp}\lesssim \lambda_1^{-2}$ can be described as independent for experimentally relevant temperatures. In case of large concentration $n_{\rm imp}\gg \lambda_1^{-2}$ we expect spin glass behavior at low temperatures due to complicated structure of the indirect exchange interaction with interactions of different sign.

We remind that a standard magnetic impurity for CdTe and HgTe is manganese cation with spin $S=5/2$ [\onlinecite{LyapilinTsidilkovski}]. Due to the presence of underlying strong spin-orbit interaction, the polarization operator at coinciding points $\Pi_{\beta\gamma,\delta\alpha}(\bm{R_A},\bm{R_A})$ has non-trivial matrix structure which results in the following form of the single-spin anisotropy:
\begin{equation}
H_{\rm anis} = V^A_{zz} S_z^{A}S_z^{A} + V^A_{xy} (S^A_x S^A_y +S^A_y S^A_x) .
\label{eq:s-anis}
\end{equation}
The last term in the right hand side of Eq. \eqref{eq:s-anis} appears due to  the presence of non-zero inversion asymmetry splitting ($V_{xy}^A=0$ for $\Delta=0$) and is also proportional to $J^A_0$.  For $V_{xy}^A=0$, Hamiltonian \eqref{eq:s-anis} describes either easy axis ($V_{zz}^A<0$) or easy plane ($V_{zz}^A>0$) anisotropy. In what follows we consider this electron-induced anisotropy \eqref{eq:s-anis} to be the main source of the anisotropy, ignoring other contributions, e.g. a strain induced anisotropy. Then in the most typical case $S=5/2$ the six degenerate levels are split into three doublets with energies $E^A_{1/2}$, $E^A_{3/2}$ and $E^A_{5/2}$ for $S_z=\pm 1/2, \pm 3/2$, and $\pm 5/2$, respectively (see Appendix \ref{App2}). It is impossible to compute accurately $V^A_{zz}$ and $V^A_{xy}$ within the Hamiltonian $\mathcal{H}$ derived in the long-wave limit. However, we can roughly estimate $|V^A_{zz}|$ and $|V^A_{xy}|$ to be of the order of $1\div 10$ K and $0.01 \div 0.1$ K, respectively. These estimates imply that the following inequality holds $|V_{zz}| \gg |V_{xy}| \gg T_*$. We also note that $V^A_{zz}$ is $10^2 \div 10^3$ times larger than the superhyperfine splitting of manganese in CdTe [\onlinecite{Lambe}].

\begin{figure*}[t]
\centerline{\includegraphics[width=0.39\textwidth]{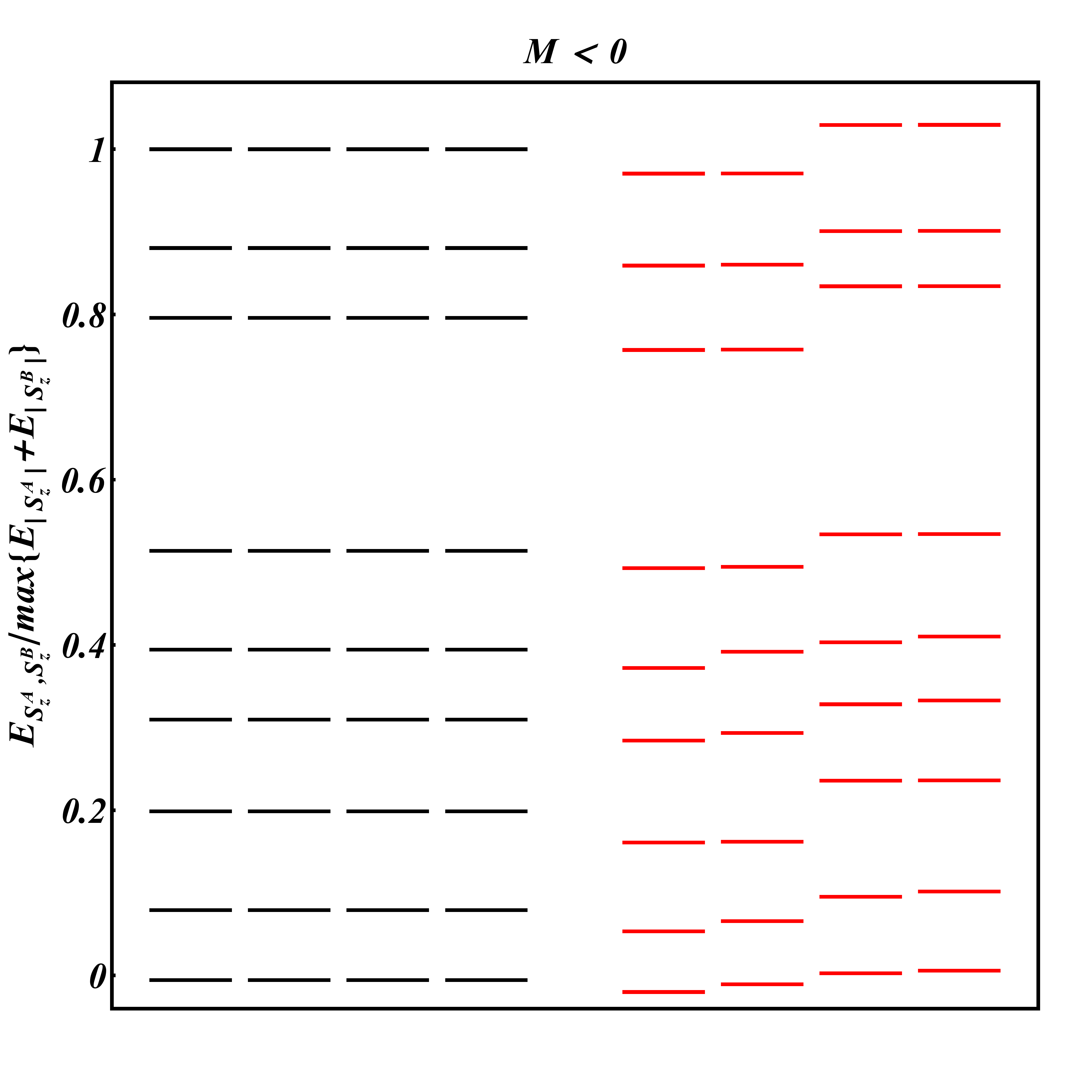}\qquad \includegraphics[width=0.39\textwidth]{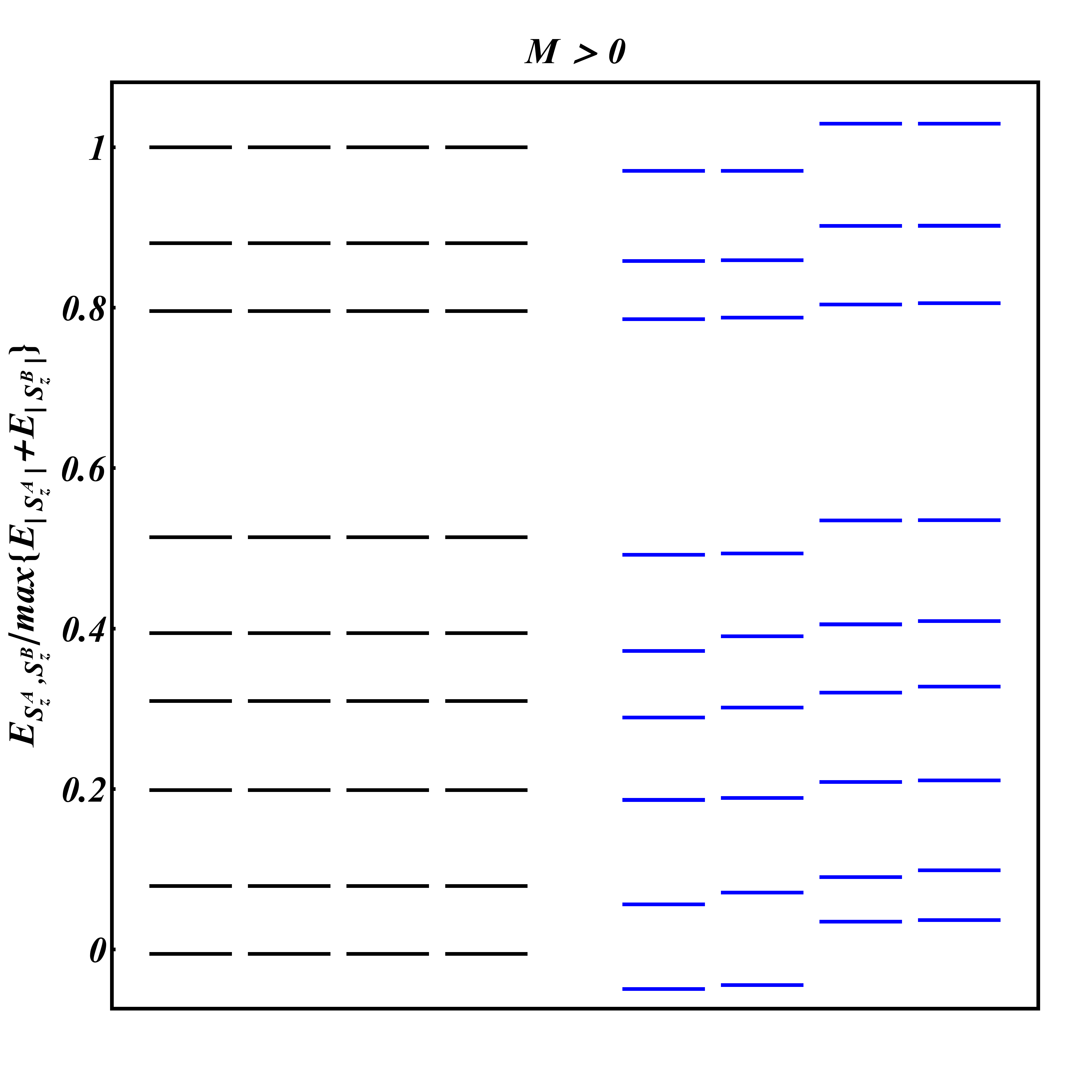}}
\caption{(Color online) The energy levels $E_{S_z^A,S_z^B}$ for the Hamiltonian $H_{\rm anis}^A+H_{\rm anis}^B+H_{IEI}$ of the two impurity problem. Splitting of 9 quartets for $M>0$ ($M<0$) is shown by blue (red) lines on the left (right) figure. The parameters are chosen as follows  $V_{zz}^A=300$, $V_{xy}^A=50$, $V_{zz}^B=75$, $V_{xy}^B=15$, $J_0^A=1$, $J_1^A=1$, $J_2^A=1$, $J_0^B=4$, $J_1^B=4$, $J_2^B=4$, $n_x = 0$, $n_y=1$, $F=1$, and $F_c=F_s=1/\sqrt{2}$. The energy levels are normalized on $\max\{E_{|S_z^A|}^A+E_{|S_z^B|}^B\}$.}
\label{Figure1}
\end{figure*}

To illustrate interesting physics of the indirect exchange interaction \eqref{eq:IEI-1} we consider how the energy levels of two manganese impurities situated at a distance $R\gtrsim  \lambda_1$ are changed due to their interaction. Since $\lambda_1\gg d$ (see Table \ref{Tab}) we assume that the energy levels $E^A_{1/2}$, $E^A_{3/2}$, $E^A_{5/2}$ and  $E^B_{1/2}$, $E^B_{3/2}$, $E^B_{5/2}$ for two impurities are different. Then without  IEI the energy spectrum of two impurities consists of 9 quartet states. We denote the corresponding energies as $E_{S_z^A,S_z^B}=E_{|S_z^A|}^A+E_{|S_z^B|}^B$, where $S_z^{A/B}=\pm 1/2, \pm 3/2$ and $\pm 5/2$. The indirect exchange interaction removes completely the degeneracy in each quartet state as shown in Fig. \ref{Figure1}. Since some terms in the interaction matrix $K_{ab}$ are proportional to $\sgn M$, the energy levels of two manganese impurities will be different for trivial insulator, $M>0$, and for topological insulator, $M<0$ (see Fig. \ref{Figure1}). As it was shown recently such fine structure of the energy levels of two magnetic impurities  can be experimentally probed by broadband electron spin resonance technique coupled with an optical detection scheme [\onlinecite{Laplane2016}].

At last, we mention that our results are also applicable to InAs/GaSb QW which is another 2D system possessing the QHS effect [\onlinecite{Liu2008},\onlinecite{Knez2011}]. Unfortunately, we cannot estimate  $\lambda_1$ and $\lambda_2$ in this case since we are not aware of detailed analysis of  parameters of the BHZ Hamiltonian for InAs/GaSb QWs.

To summarize, we studied the indirect exchange interaction between magnetic impurities mediated by virtual interband transitions of 2D quasiparticles in the CdTe/HgTe/CdTe QW at low temperatures. We found
the exponential decay with the distance of the indirect exchange interaction. In addition to the exponential decay of the indirect exchange interaction we obtained its oscillations with the distance due to the presence of inversion asymmetry in the QW. In general, the interaction matrix $K_{ab}$ has complicated structure with dependence on the unit vector along the direction between two impurities (see Eq. \eqref{eq:IEI-2}). In the absence of inversion asymmetry the indirect exchange interaction involves anisotropic XXZ Heisenberg interaction, magnetic pseudo-dipole interaction, and Dzyaloshinsky-Moriya interaction. The sign of Dzyaloshinsky-Moriya interaction depends on the sign of the bulk gap $M$. 

During the preparation of the manuscript we became aware of the work by Litvinov [\onlinecite{Litvinov2016}]
where IEI between magnetic impurities at the surface of a thin film of 3D TI was studied at zero temperature. It was found that for the chemical potential of surface states lying within the gap an exponential decay of IEI between magnetic impurities polarized perpendicular to the surface is accompanied by oscillations with the spatial period of a few nanometers. The behavior of IEI predicted in Ref. [\onlinecite{Litvinov2016}] is qualitatively similar to the dependence of $K_{zz}$ on $R$ which we reported in this paper (see Eq. \eqref{eq:IEI-2}).

\begin{acknowledgments}

We thank B. Aronson, M. Durnev, M. Feigel'man, M. Glazov,  G. Min'kov, I. Rozhansky and, especially, S. Tarasenko, for useful discussions. The work was partially supported by the Russian Foundation for Basic Research under the Grant No. 15-52-06005, Russian President Grant No. MD-5620.2016.2, and the Ministry of Education and Science of the Russian Federation under the Grant No. 14.Y26.31.0007.

\end{acknowledgments}

\appendix

\begin{widetext}
\section{Evaluation of $\Pi_{\beta\gamma,\delta\alpha}$\label{App1}}

In this Appendix we present the explicit procedure of the evaluation of the polarization operator $\Pi_{\beta\gamma,\delta\alpha}$. We consider a more general form of the BHZ Hamiltonian. In the dimensionless variables it can be written as 
\begin{equation} 
\frac{H(\mathbf{\bm\kappa})}{|m|{\cal{E}}}= -d(\kappa^2) + \begin{pmatrix}b(\kappa^2)\sgn m &{\cal{A}}(\kappa^2)\kappa_+&0&\gamma(\kappa^2)\\ 
{\cal{A}}(\kappa^2)\kappa_-&-b(\kappa^2)\sgn m &-\gamma(\kappa^2)&0\\ 
0&-\gamma(\kappa^2)&b(\kappa^2)\sgn m &-{\cal{A}}(\kappa^2)\kappa_-\\ 
\gamma(\kappa^2)&0&-{\cal{A}}(\kappa^2)\kappa_+&-b(\kappa^2)\sgn m \\ 
\end{pmatrix} 
\label{App1:H}
\end{equation} 
where $b(0)=1$, $b(\kappa \sim 1)-b(0) \sim |m|$, $d(\kappa \sim 1) \sim |m|$, ${\cal{A}}(0)=1$, $\gamma(0) \equiv \gamma \sim 1$, $\gamma(\kappa \sim 1) \sim 1$. 
Some additional restrictions of the model Hamiltonian \eqref{App1:H} will be discussed later.

We start from the derivation of Eq. \eqref{eq:Pi}. The polarization operator \eqref{H-IEI} can be explicitly written at $T=0$ as
\begin{gather} 
\Pi_{\beta\gamma,\delta\alpha}=\sum_{j,l=1,\dots,4}\frac{|m|^4}{a^4}\int \frac{d\varepsilon}{2\pi}\int \frac{d^2\bm{\kappa}_1d^2\bm{\kappa}_2}{(2\pi)^4}\mathcal{R}_{\beta j}(\bm{\kappa}_1) \hat{\mathcal{G}}_{jj}(i\varepsilon,\bm{\kappa}_1) \mathcal{R}_{j\gamma}^{-1}(\bm{\kappa}_1) \mathcal{R}_{\delta l}(\bm{\kappa}_2) \hat{\mathcal{G}}_{ll}(i\varepsilon,\bm{\kappa}_2) 
\mathcal{R}^{-1}_{l \alpha}(\bm{\kappa}_2)e^{i(\bm{\kappa}_1-\bm{\kappa}_2)\bm{\rho}} ,
\label{App1:Pi}
\end{gather} 
where $\hat{\mathcal{G}}$ stands for the Green's function in the eigen basis of the Hamiltonian \eqref{App1:H} (see Eq. \eqref{eq:Gdiag}).

We notice that in order for the integral over $\varepsilon$ in Eq. \eqref{App1:Pi} to be non-zero, the poles must lay on the different sides from the chemical potential. Integrating over the poles of the Green's functions and changing $\bm \kappa_1 \rightarrow -\bm \kappa_2$, $\bm \kappa_2 \rightarrow -\bm \kappa_1$ in some of the terms (this is important to get common denominator, leaving the exponent the same) we get: 
\begin{gather}
\Pi_{\beta\gamma,\delta\alpha}=-\frac{|m|^3}{a^4\mathcal{E}}\sum_{j=1,3}\sum_{l=2,4}\int\frac{d^2 \kappa_1}{(2\pi)^2}\frac{d^2 \kappa_2}{(2\pi)^2}e^{i\bm \rho (\bm \kappa_1 - \bm \kappa_2)}\frac{1}{\mathrm{sgn}M(\epsilon_j(\bm \kappa_1)-\epsilon_l(\bm \kappa_2))}\\
\notag \times \left( \mathcal{R}_{\beta j}(\bm \kappa_1)\mathcal{R}_{j \gamma}^{-1}(\bm \kappa_1)\mathcal{R}_{\delta l}(\bm \kappa_2)\mathcal{R}^{-1}_{l\alpha}(\bm \kappa_2) +  \mathcal{R}_{\beta l}(-\bm \kappa_2)\mathcal{R}_{l \gamma}^{-1}(-\bm \kappa_2)\mathcal{R}_{\delta j}(-\bm \kappa_1)\mathcal{R}^{-1}_{j\alpha}(-\bm \kappa_1)\right) .
\end{gather}
Let's, as previously, introduce $\mathcal{X}_\pm=\sqrt{({\cal{A}}\kappa\pm \gamma)^2+b^2}$. Grouping various terms and using the relation $1/r=\int_0^\infty dt \exp(-rt)$, we obtain Eq. \eqref{eq:Pi}.

It is convenient to introduce the following notations
\begin{equation}
\begin{split}
L^\pm_\nu & =\int\limits_{-\infty}^\infty d\kappa\frac{|\kappa|}{{\cal{X}}_-}e^{-t {\cal{X}}_-}e^{\sgn m t d(\kappa^2)}J_\nu (\kappa \rho)({\cal{X}}_-\pm b(\kappa^2)), \\
  P^\pm_\nu & =\int\limits_{-\infty}^\infty d\kappa\frac{|\kappa|}{{\cal{X}}_-}e^{-t {\cal{X}}_-}e^{-\sgn m td(\kappa^2)}J_\nu (\kappa \rho)({\cal{X}}_-\pm b(\kappa^2)), \\
L_\nu & =\int\limits_{-\infty}^\infty d\kappa\frac{|\kappa|}{{\cal{X}}_-}e^{-t {\cal{X}}_-}e^{\sgn m td(\kappa^2)}J_\nu (\kappa \rho)(\gamma(\kappa^2)-{\cal{A}}(\kappa^2)\kappa), \\
P_\nu & = \int\limits_{-\infty}^\infty d\kappa\frac{|\kappa|}{{\cal{X}}_-}e^{-t {\cal{X}}_-}e^{-\sgn m td(\kappa^2)}J_\nu (\kappa \rho)(\gamma(\kappa^2)-{\cal{A}}(\kappa^2)\kappa)  ,
\end{split}
\label{App1:LP}
\end{equation}
where $J_\nu (\kappa \rho)$ stands for the Bessel function.
Then the direct evaluation leads to Eq. \eqref{eq:IEI-1} with the following interaction matrix
\begin{equation} \label{App1:eq:K-func} 
K_{ab}=\frac{|m|^3}{16 \pi^2a^4 {\cal{E}} } \int\limits_0^{\infty} dt \, U_{ab}(\rho,t) 
\end{equation}
where
\begin{equation}
\begin{split} 
U_{xx} = & -J_0^A J_0^B\Bigl [2(2 P_1 L_1-P_1^- L_1^- -P_1^+ L_1^+)n_x^2-(2 P_0 L_0+2 P_1 L_1-P_0^- L_0^- -P_0^+ L_0^+-P_1^- L_1^- -P_1^+ L_1^+)\Bigr ]\\
&-J^A_{m} J^B_{m}\Bigl [ P_0^- L_0^+ +P_1^- L_1^+(n_x^2-n_y^2)\Bigr ]-2(J^A_{m}J^B_0+J^A_0 J^B_{m}) (L_1 P_1^- - P_1 L_1 ^+)n_xn_y  ,   \\
U_{xy}=& -2J_0^A J_0^B(2 P_1 L_1-P_1^- L_1^- -P_1^+ L_1^+)n_x n_y+2J^A_{m} J^B_{m}P_1^- L_1^+n_x n_y\\
&+\sgn m (J^A_{m} J^B_0 + J^A_0 J^B_{m})(L_0 P_0^- - P_0 L_0 ^+)+\sgn m (J^A_{m}J^B_0-J^A_0 J^B_{m}) (L_1 P_1^- - P_1 L_1 ^+)(n_x^2-n_y^2) ,   \\
U_{zz} =& -\Bigl [ (J_2^A J_1^B+J_1^A J_2^B)(P_0 L_0-P_1 L_1)+J^A_2 J^B_2(L_0^- P_0^+-L_1^- P_1^+)+J^A_1J^B_1(P_0^- L_0^+-P_1^- L_1^+)\Bigr ] ,  \\
U_{xz} =& \sgn m J^A_0 \Bigl [ J^B_1(P_0^- L_1+P_1^- L_0-P_1 L_0^+-P_0 L_1^+)+J_2^B(P_1 L_0^- +P_0 L_1^- - P_0^+ L_1 -P_1^+ L_0)\Bigr ] n_x+\\
&-J^A_{m}\Bigl [ J^B_2(P_1 L_0 +P_0 L_1)+J^B_1(P_1^- L_0^+ +P_0^- L_1^+) \Bigr] n_y  .
\end{split}
\label{App1:UU}
\end{equation}
The elements $U_{yy}$, $U_{yx}$ and $U_{yz}$ can be found from $U_{xx}$, $U_{xy}$ and $U_{xz}$ by interchange of $n_x$ and $n_y$. The elements $U_{zx}$ and $U_{zy}$ are equal to  $-U_{xz}$ and $-U_{yz}$ with superscripts $A$ and $B$ interchanged. The structure of the indirect exchange interaction is given by Eq. \eqref{App1:UU}. Below we investigate asymptotic behavior of the indirect exchange interaction at long distances where it is determined by the universal long distance part of the Hamiltonian \eqref{App1:H}. So we need to analyze all the integrals listed in Eq. \eqref{App1:LP}. 

Let us start from the integral $L_0^+$ with $m>0$ as an example. Asymptotic expressions for all the other integrals are calculated in a similar way. So we wish to compute
\begin{equation}\label{App-2}
L_0^+=\int\limits_{-\infty}^\infty d\kappa \frac{|\kappa|}{{\cal{X}_-}}e^{td(\kappa^2)-t{\cal{X}_-}}J_0(\kappa \rho)({\cal{X}_-}+b(\kappa^2)) .
\end{equation}
The following problem arises: integrand is not an analytical function of $\kappa$ because of the presence of $|\kappa|$. In order  to work with this function we represent $|\kappa|$ as  $|\kappa|_q\equiv \sqrt{\kappa^2-q^2}$ where the limit $q\to 0$ is assumed. The cuts are chosen to go along the real axis from $-\infty$ to $-q$ and from $q$ to $+\infty$. For this branch choice one can see that in the upper semiplane $|\kappa|_q=\kappa$ and $|\kappa|_q=-\kappa$ in the lower semiplane. The contour of integration is slightly deformed so that it goes above the cut for $\kappa>0$ and below the cut for $\kappa<0$ (see Fig. \ref{Figure-k-plane-1}).
Next we use the following relation which is valid on the real axis:
$J_0(z)=\re H^{(1)}_0(|z|)$. This substitution is motivated by the exponential growth of the Bessel function $J$ on the imaginary axis. The Hankel function $H^{(1)}_0$ decays exponentially along the imaginary axis. 

Now some restrictions must be imposed on $\mathcal{X}_-$. We assume that $\mathcal{X}_-$ is a square root of a polynomial of a finite degree $\mathcal{X}_-=\sqrt{f(\kappa)}$. This is correct provided $b$, $\mathcal{A}$, and $\gamma$ are also polynomial in $\kappa$. We will also suppose that the polynomial $f(\kappa)$ has only one minima at the point $\kappa_0\sim \gamma$ and, moreover, $\mathcal{X}_-\pm d$ also has one minima. We emphasize that all these requirements correspond to the case of the BHZ Hamiltonian, which is the main target of our consideration.

Taking  into account all the remarks above, we obtain
\begin{equation}
L_0^+=\mathrm{Re}\int\limits_{-\infty}^\infty d\kappa\frac{|\kappa|_q}{\sqrt{f(\kappa)}}e^{-t \sqrt{f(\kappa)}}e^{t d(\kappa^2)}H^{(1)}_0 (|\kappa|_q\rho)\bigl (\sqrt{f(\kappa)}+b(\kappa^2)\bigr ) .
\label{App1:L01}
\end{equation}
We divide the region of the integration into two parts: integration from $-\infty$ to the minima at $\kappa_0$ of $f(\kappa)$, and from $\kappa_0$ to $+\infty$. Within the intervals $\kappa<\kappa_0$ and $\kappa>\kappa_0$ the function $f(\kappa)$ is monotonous, therefore it is convenient to use variable  $\epsilon=\sqrt{f(\kappa)}$ instead of $\kappa$. Then Eq. \eqref{App1:L01} can be rewritten as
\begin{equation}
L_0^+= \re\Biggl\{ -\int\limits_{\epsilon_{0}}^{\infty}d\epsilon \frac{|\kappa_{1}|_q}{\epsilon}\frac{d\kappa_{1}}{d\epsilon}e^{t d(\kappa_1^2)}e^{-t\epsilon}H_{0}^{(1)}(|\kappa_{1}|_q\rho)\bigl (\epsilon+b(\kappa_1^2)\bigr )+\int\limits_{\epsilon_{0}}^{\infty}d\epsilon \frac{|\kappa_{2}|_q}{\epsilon}\frac{d\kappa_{2}}{d\epsilon}e^{t d(\kappa_2^2)}e^{-t\epsilon}H_{0}^{(1)}(|\kappa_{2}|_q\rho)\bigl (\epsilon+b(\kappa_{2}^{2})\bigr )\Biggr \} .
\label{App1:L02}
\end{equation}
\end{widetext}
Here $\epsilon_0=\sqrt{f(\kappa_0)}$ and $\kappa_{1,2}$ are two real solutions of the equation $\sqrt{f(\kappa)}=\epsilon$ for $\epsilon>\epsilon_0$ (see Fig. \ref{Figure2}). The integrals in the expression above can be reduced to a single integration in the complex plane of $\epsilon$.
In the vicinity of the minima, the energy can be written as $\epsilon(\kappa)=\epsilon_0+\epsilon^{\prime \prime}_0 (\kappa-\kappa_0)^2/2+\dots$ and thus the inverse function $\kappa(\epsilon)=\kappa_0 \pm \sqrt{2(\epsilon-\epsilon_0)/\epsilon^{\prime \prime}_0}+\dots$ has a branch cut. We choose the branch cut to go from $\epsilon_0$ along the real axis to the infinity. Then we choose the branch of the square root in such a way that $k(\epsilon)\equiv k_{1/2}(\epsilon)$ above/below the real axis. As a result
\begin{equation}
L_0^+= \re \int\limits_{C} d\epsilon \frac{|\kappa|_q}{\epsilon}\frac{d\kappa}{d\epsilon}e^{t d(\kappa^2)-t\epsilon}H_{0}^{(1)}(|\kappa|_q\rho)\bigl (\epsilon+b(\kappa^2)\bigr ) ,
\label{App1:contour}
\end{equation}
where the contour $C$ is depicted in the Fig. \ref{Figure-energy}.

Now in order to estimate the integrals for large values of $\rho$ and $t$ the steepest descent method can be used. As we will see below at the saddle point the following inequality holds: $|\kappa|\rho\gg 1$. Therefore we can use the asymptotic expansion for the Hankel function:
\begin{equation}
H_0^{(1)}(x)\approx\sqrt{{2}/({\pi x})}\exp( i x - i{\pi}/{4}) .
\end{equation}

The saddle point equation for the integral reads
\begin{equation}
\frac{d\kappa}{d\epsilon} = -i\frac{t}{\rho}\left (1-\frac{d\, d(\kappa^2)}{d\kappa}\frac{d\kappa}{d\epsilon}\right ) .
\end{equation}
This is equivalent to the following 
\begin{equation}
\epsilon =  -i f^\prime(\kappa(\epsilon))\frac{t}{2\rho}\left (1-\frac{d \, d(\kappa^2)}{d\kappa}\frac{d\kappa}{d\epsilon}\right ) .
\end{equation}
As we will demonstrate below values of $t$ which dominate the integral in Eq. \eqref{App1:eq:K-func} satisfy inequality $t\ll \rho$.
Using ${t}/{\rho}\ll 1$ as a small parameter, we find the leading order solution of the saddle point equation: 
\begin{gather}
\epsilon= -i \frac{t}{2\rho}f^\prime(\kappa(0)) .
\end{gather}
The second derivative of $\kappa$ at the saddle point  is given to the leading order as:
\begin{gather}
\frac{d^{2}\kappa}{d\epsilon^{2}}=\frac{2}{f^\prime(\kappa(0))} \, .
\end{gather}
\begin{figure}[t] 
\centerline{\includegraphics[width=0.4\textwidth]{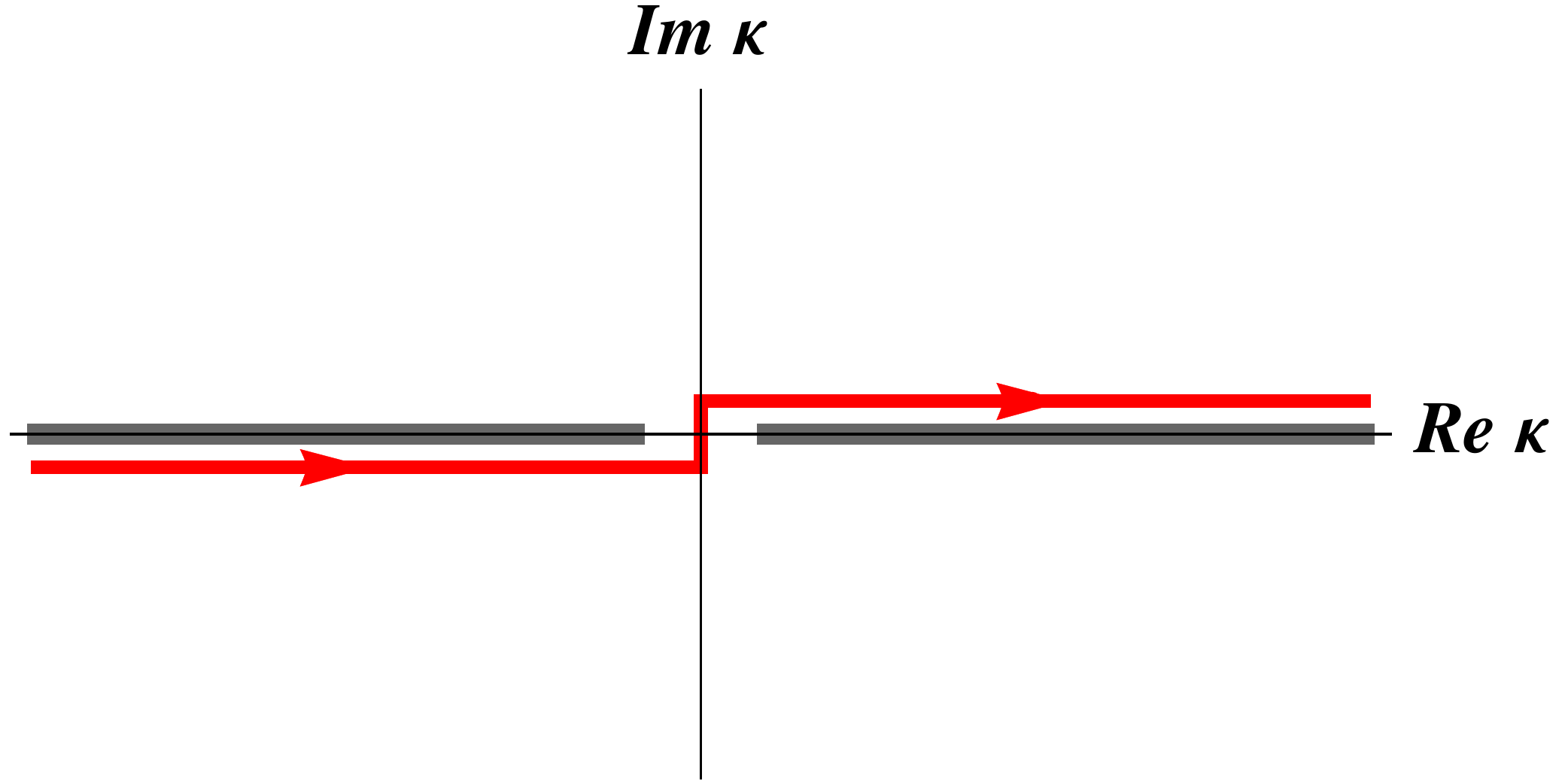}} 
\caption{(Color online) The contour of integration in the $\kappa$ plane. The branch cuts of $|\kappa|_q$ are shown by thick gray lines.}
\label{Figure-k-plane-1} 
\end{figure}

In order to apply the steepest decent method we need to deform the contour of integration in the $\epsilon$ plane  in such a way so that passes through the saddle point. The convenient contour is $\widetilde{C}$ shown in Fig. \ref{Figure-energy}. However, deforming the contour in the $\epsilon$ plane in this way we have to pass the branching point of $|\kappa(\epsilon)|_q$. Further we proceed as if there is no such problem and discuss how to overcome it later. As we will demonstrate it does not affect the final result.

Now we define convenient notations
\begin{equation} 
\begin{split}
f^\prime(\kappa(0))=i c^2_f e^{2 i \phi_f},\\
 \kappa(0)= i\xi e^{i \phi_{\xi}},\\
 b(\kappa^2(0))=c_b e^{i \phi_b}  .
\end{split}
\label{App1:Not}
\end{equation}
We note that non-zero $\phi_f$ corresponds to the existence of the imaginary part of the saddle point. Choosing the contour such that it goes along the steepest descent direction and evaluating the Gaussian integral, we find
\begin{align} 
\label{App1:eq: result} 
L_0^+=& \re \Biggl [ \frac{2}{c_{f}}\sqrt{\frac{2\xi}{\rho^{2}}}c_b e^{i \phi_b}e^{t d(\kappa_1^2(0))}
\exp\left (-\frac{t^{2}}{4\rho}c_{f}^{2}e^{2i\phi_{f}}\right ) \notag \\
& \times \exp \left ( -\xi e^{i\phi_{\xi}}\rho-i\phi_{f}+\frac{i\phi_{\xi}}{2}\right )\Biggr ] .
\end{align} 
Now we discuss several issues which were important for the evaluation of the integral above.  In Eq. \eqref{App1:eq: result} there is a term proportional to $t^2/\rho$ in the exponent. Since $c_f\sim 1$ the integration over $t$ in Eq. \eqref{App1:eq:K-func} will be dominated by $t \lesssim\sqrt{\rho}$, i.e. $t/\rho \lesssim 1/\sqrt{\rho} \ll 1$ for $\rho \gg 1$ which was necessary for the perturabtive expansion of the saddle point.

Let us discuss the question about the branching of $|\kappa|_q$. The point is that the contour $C$ that surrounds the cut in $\epsilon$ plane cannot be simply deformed to the straight line ($\widetilde{C}$) that passes through the saddle point because there is an additional cut due to $|\kappa|_q=\sqrt{\kappa^2-q^2}$. In fact, this difficulty can be overcome as follows. For this purpose, it is more convenient to track deformations of the contour in the $\kappa$ plane. The contour $\widetilde{C}$ near the saddle point in the $\epsilon$ plane corresponds to the contour $C_0$ in the $\kappa$ plane (see Fig. \ref{Figure-k-plane-2}). Such deformation of the contour depicted in Fig. \ref{Figure-k-plane-1} implies necessarily one more contour ($C_1$) which surrounds the negative real semi axis (see Fig. \ref{Figure-k-plane-2}). However, the real part of the integral along the contour $C_1$ is zero. This happens because (i) the function $\re \left[ |\kappa|_q H_0^{(1)}(|\kappa|_q \rho )\right]$ is exactly the same for $\kappa = -\delta \pm  i0$ with $\delta > 0$, (ii) the directions of integration below and above the negative semi axis are opposite to each other.
\begin{figure}[t] 
\centerline{\includegraphics[width=0.4\textwidth]{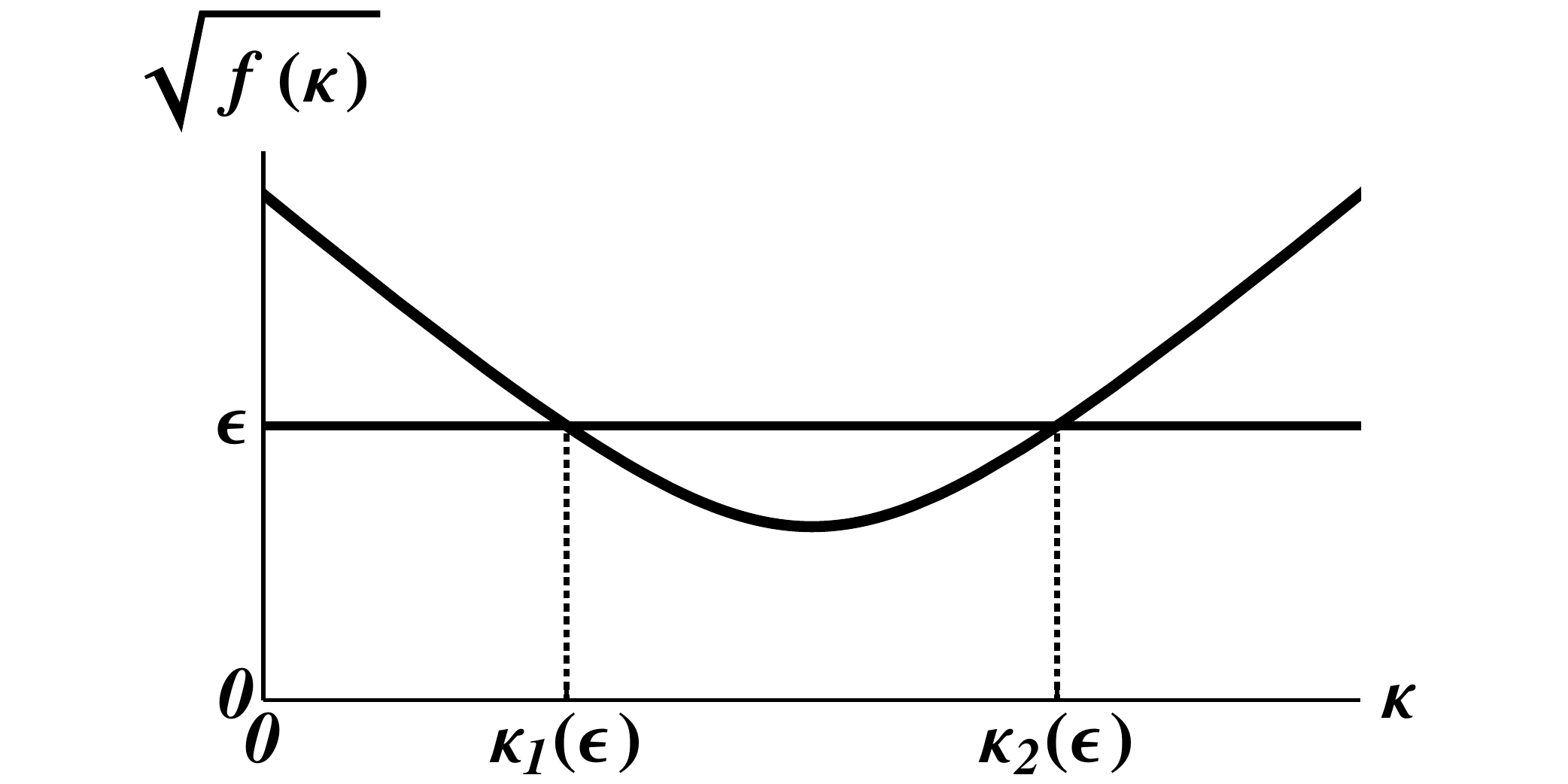}} 
\caption{Solutions of the equation $\epsilon=\sqrt{f(\kappa)}$ for $\epsilon>\epsilon_0$.}
\label{Figure2} 
\end{figure}

In the saddle point approximation we can omit the term $\pm t d(\kappa^2)$ in the exponents of the $L$ and $P$ integrals. Indeed, the quantities $U_{ab}$ in Eq. \eqref{App1:UU} involve products of the $L$ and $P$ integrals. Therefore the real part of $d(\kappa_1^2(0))$ does not contribute to $U_{ab}$. 
Provided $\rho \ll {1}/{(|m|^2\gamma^2)}$ we can neglect the imaginary part of $d(\kappa_1^2(0))$. We note that  for the BHZ model $1/(|m|^2\gamma^2)$ can be estimated to be of the order of $10^4$. In what follows we will not make distinction between $L$ and $P$ integrals. 

Within the saddle point approximation the $L$ integrals can be expressed in terms of $L_0^+$. Since one finds $\mathcal{X}_- = 0$ in the lowest order in the parameter $t/\rho$,  we obtain $L^+_\nu=-L^-_\nu$. 
The expressions for $L_0$ can be obtained by a $\pi/2$ phase shift of $\phi_b$ from the result for $L_0^+$, 
because at the saddle point the following relation holds: $\gamma(\kappa^2(0))-{\cal{A}}(\kappa^2(0))\kappa(0)=-i b(\kappa^2(0))$. Using for $x\gg 1$ the asymptotic relation $H_1(x) \simeq -i H_0(x)$, we  express the integrals $L_1^+$ and $L_1$ in the following way: 
$L_1^+=L_0$ and $L_1=-L_0^+$. 

\begin{figure}[t]
\centerline{\includegraphics[width=0.4\textwidth]{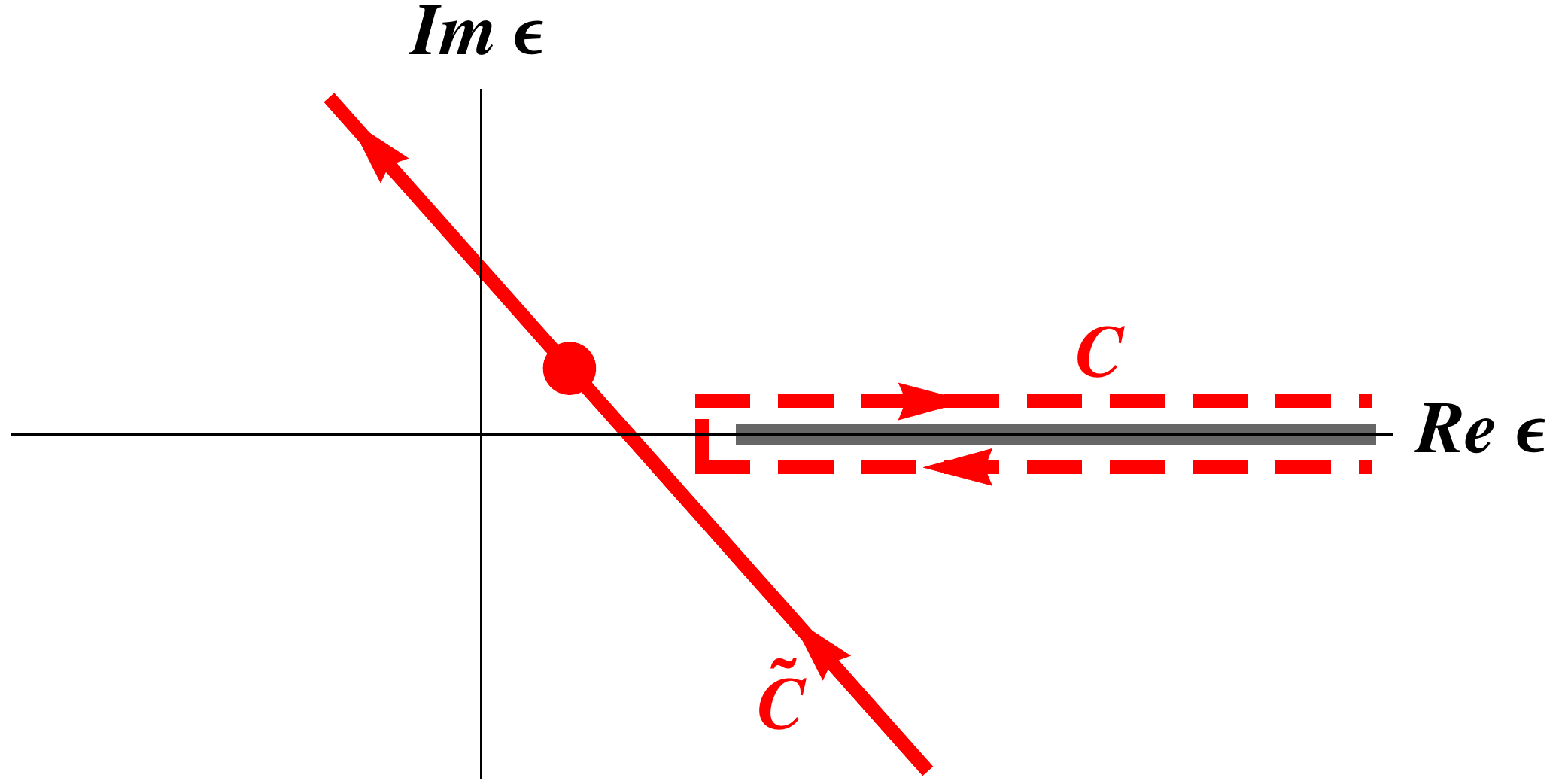}}
 \caption{(Color online) The contours of integration in the $\epsilon$ plane before ($C$) and after ($\widetilde{C}$) deformation. Non-zero $\phi_f$ (see Eq. \eqref{App1:Not}) implies that the saddle-point is situated away from the real axis.}
 \label{Figure-energy}
\end{figure}

Introducing the following functions
\begin{align}
F_c(R) & =\frac{|m|^3}{16 \pi^2a^4 \mathcal{E}}\int_{0}^{\infty}dt \left( {L^+_0}^2-{L^+_1}^2 \right) ,\notag
\\
F_s(R)& =\frac{|m|^3}{16 \pi^2a^4\mathcal{E}}\int_{0}^{\infty}dt  \left(2 L^+_0 L^+_1 \right) , 
\\
F(R)&=\frac{|m|^3}{16\pi^2a^4\mathcal{E}}\int_{0}^{\infty}dt \left( {L^+_0}^2+{L^+_1}^2 \right) ,\notag
\end{align}
we obtain the result \eqref{eq:IEI-1} -- \eqref{eq:IEI-2}. Integrating over $t$, we find the following explicit expressions for the functions $F$, $F_s$, and $F_c$:
\begin{equation}
\begin{split}
F(R)& =\frac{|m|^3}{a^4 \mathcal{E}}\frac{\xi c_b^2}{c_f^3}\frac{e^{-2\xi\cos{\phi_\xi}\rho}}{(2\pi\rho)^{3/2}}\frac{1}{\sqrt{\cos(2\phi_f)}} ,
\\
F_c(R)& =F(R)\sqrt{\cos(2\phi_f)}\cos{\left( 2\xi \sin{\phi_\xi}\rho-\tilde{\phi}\right)} ,
\\
F_s(R) & =-F(R) \sqrt{\cos(2\phi_f)}\sin{\left( 2\xi \sin{\phi_\xi}\rho-\tilde{\phi}\right)} ,
\end{split}
\end{equation}
where $\tilde{\phi} = \phi_\xi+2\phi_b-3\phi_f$ and  $\rho = a R/|m|$. Using the following relations 
$\phi_b=\phi_f=0$, $c_f = \sqrt{2} (1+(1+\gamma^2) m \cosh\chi /2)$, $c_b = \xi \cos\phi_\xi$, and
\begin{gather}
\xi e^{i\phi_\xi} = 1-(1-\gamma^2)m\cosh\chi - i \gamma (1-2 m \cosh\chi) ,
\end{gather}
which are valid to the lowest order in $|m|$, we obtain the result \eqref{eq:IEI-3}.

\begin{figure}[t]
\centerline{\includegraphics[width=0.4\textwidth]{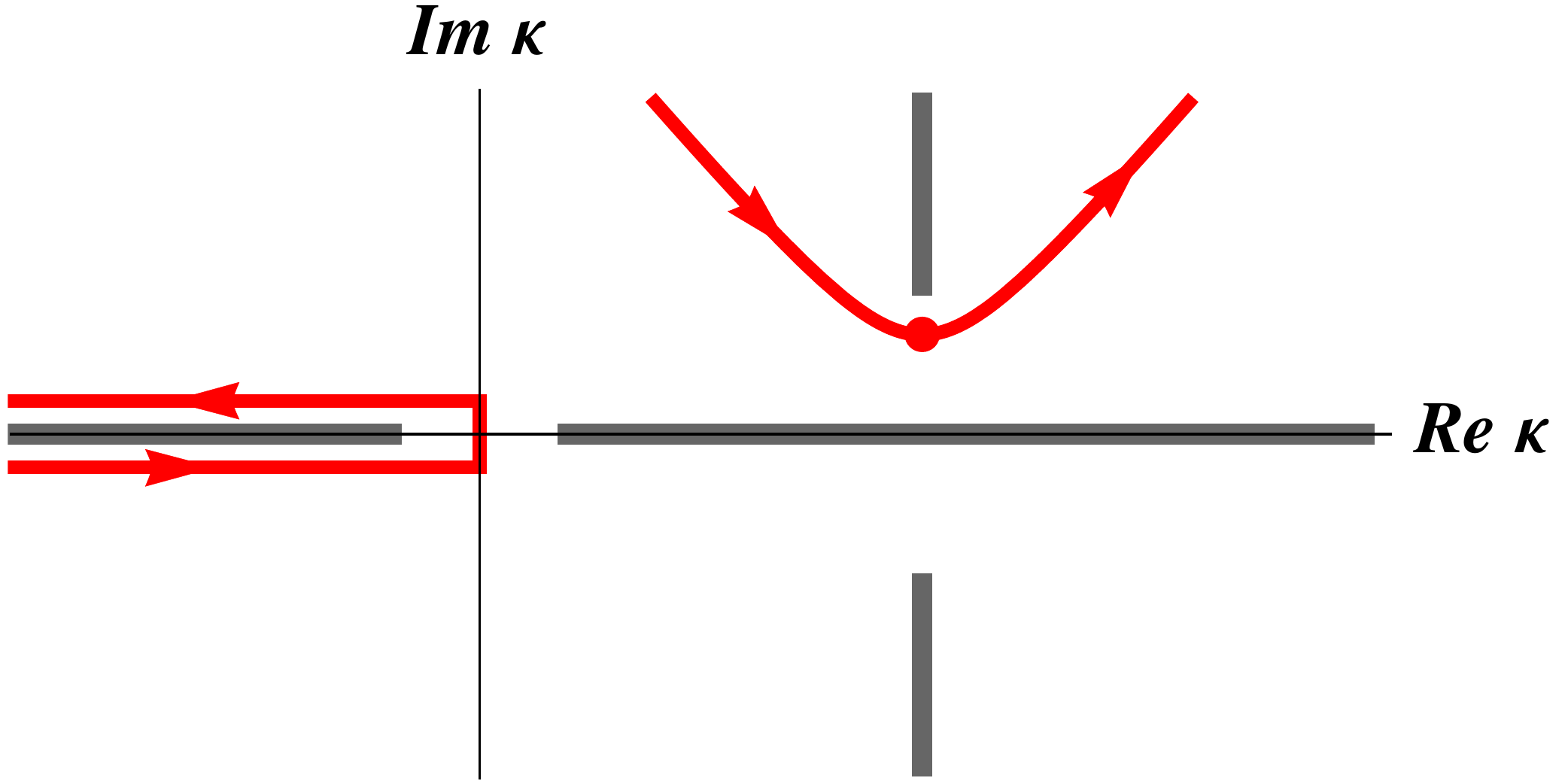}}
 \caption{(Color online) The contour of integration in $\kappa$ plane is depicted schematically by red curves. The branch cuts for $|\kappa|_q$ and $\sqrt{f(\kappa)}$ are shown by gray lines.}
 \label{Figure-k-plane-2}
 \end{figure}

\section{The single-spin anisotropy Hamiltonian \label{App2}}

We begin the analysis of the Hamiltonian \eqref{eq:s-anis} from estimates for the values of $V_{zz}$ and $V_{xy}$. For an order of magnitude estimate it is enough to evaluate the integrals in Eq. \eqref{App1:eq:K-func} at $\rho=0$ keeping in mind that the cut-off at large wave vectors (close to the inverse lattice constant which is of order $1/a$) should be introduced. This cut-off is equivalent to the cut-off of the order of $|m|$ for the integral over $t$ from below. To the lower order in $1/t$ we estimate
$U_{xx}\sim U_{zz}\sim 1/t^4+\dots$ and $U_{xy}\sim \gamma/t^3+\dots$. $U_{xz}$ turns out to be zero in this rough estimate because it involves integrals with $J_1(\kappa \rho)$ which is zero for $\rho=0$.

Evaluating the integral over $t$ one obtains 
$V_{zz}\sim 1/({\cal{E}}a^4)$ and $V_{xy}\sim |m|\gamma /({\cal{E}}a^4)$. It is worthwhile to mention that (i) 
$V_{xy}$ vanishes in the absence of interface inversion asymmetry (for $\gamma=0$), (ii) for $\gamma \sim 1$ the following estimate holds $|V_{xy}/V_{zz}|\sim |m| \sim 0.1 \div 0.01$, i.e. $|V_{zz}|\gg |V_{xy}|$.

\begin{figure}[t]
\centerline{\includegraphics[width=0.4\textwidth]{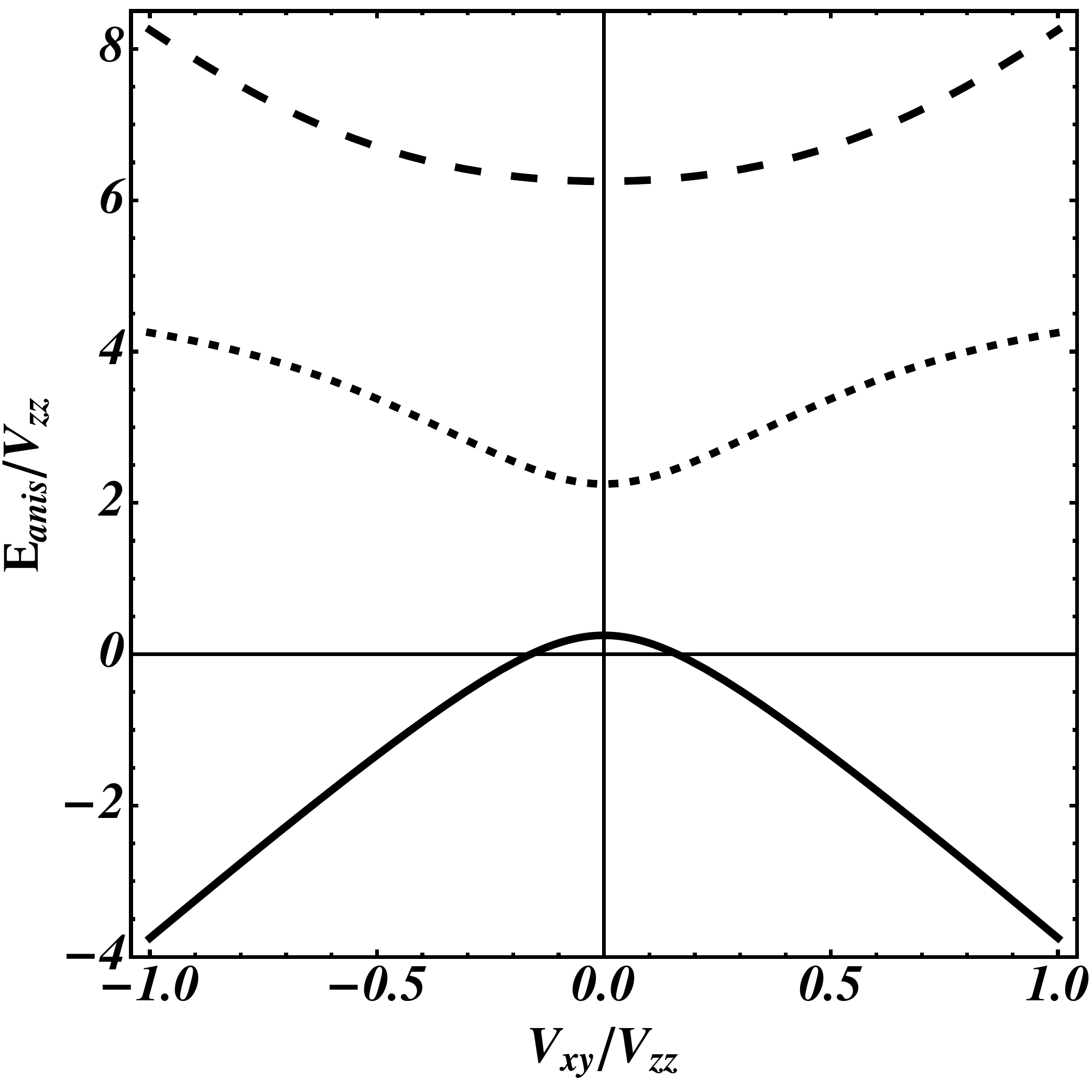}}
 \caption{Energy levels for Hamiltonian $H_{\rm anis}$ as a function of $V_{xy}/V_{zz}$. It is assumed that $V_{zz}>0$.}
 \label{Figure-anis}
\end{figure}

Now we diagonalize Hamiltonian \eqref{eq:s-anis} for $S=5/2$. It is convenient to choose the following basis 
of the states with a given spin projection on the $z$-axis: $|-5/2\rangle$, $|-1/2\rangle$, $|3/2\rangle$, $|5/2\rangle$,  $|1/2\rangle$, $|-3/2\rangle$. Then Hamiltonian \eqref{eq:s-anis}  can be written as a block-diagonal matrix $6\times 6$:
\begin{equation}
H_{\mathrm{anis}}=
\begin{pmatrix}
H_{\mathrm{anis}}^{3\times 3} & 0 \\
0 & H_{\mathrm{anis}}^{3\times 3}
\end{pmatrix}
\label{App2:a-exp-1}
\end{equation}
where
\begin{equation}
H_{\mathrm{anis}}^{3\times 3}=
\begin{pmatrix}
\frac{25}{4}V_{zz} & -i\sqrt{10}V_{xy} & 0\\
i\sqrt{10}V_{xy} & \frac{1}{4}V_{zz} & -3i\sqrt{2}V_{xy}\\
0 & 3i\sqrt{2}V_{xy} & \frac{9}{4}V_{zz}
\end{pmatrix} .
\label{App2:a-exp-2}
\end{equation}
It means the for an arbitrary ratio $V_{xy}/V_{zz}$ there are three twofold degenerate eigenvalues of $H_{\mathrm{anis}}$. Solving the corresponding cubic equation, one can find these eigenvalues exactly. The full result is bulky and thus omitted here, the eigenvalues for the different ratios between $V_{xy}$ and $V_{zz}$ are shown in the Fig. \ref{Figure-anis}.

Since according to the estimation above $|V_{xy}| \ll |V_{zz}|$, one can treat the second term in the right hand side of Eq. \eqref{eq:s-anis} as the perturbation. For $V_{xy}=0$ the eigenvalues of the latter are $E^{(0)}_{1/2}=V_{zz}/4$, $E^{(0)}_{3/2}=9V_{zz}/4$ and $E^{(0)}_{5/2}=25V_{zz}/4$. They don't depend on the sign of $S_z$. Once the perturbation is added $S_z$ no longer commutes with $H_{\mathrm{anis}}$ and then the projection of the impurity's spin on the $z$-axis is not a good quantum number anymore. However, for convenience we will denote the eigenvalues of $H_{\mathrm{anis}}^{3\times 3}$ as $E_{1/2}$, $E_{3/2}$ and $E_{5/2}$. The standard second order perturbation theory leads to the following results
\begin{align}
E_{1/2} & = \frac{1}{4}V_{zz}-\frac{32}{3}\frac{V_{xy}^2}{V_{zz}} ,\notag
\\
E_{3/2} & = \frac{9}{4}V_{zz}+9\frac{V_{xy}^2}{V_{zz}} ,
\\
E_{5/2} & = \frac{25}{4}V_{zz}+\frac{5}{3}\frac{V_{xy}^2}{V_{zz}} .\notag
\end{align}\label{App2:lvls-approx}

\bibliography{biblio.bib}

\end{document}